\documentclass[english,letterpaper,twocolumn,aps,prl,showpacs]{revtex4}
\usepackage[T1]{fontenc}
\usepackage[latin1]{inputenc}
\usepackage{graphicx}
\usepackage{amssymb}

\makeatletter

\makeatletter

\makeatletter

\makeatletter

\makeatletter

\input{epsf}

\makeatother

\makeatother

\makeatother

\makeatother

\usepackage{babel}
\makeatother
\begin{document}

\title{Visualization of vortex bound states in polarized Fermi gases at
unitarity}

\author{Hui Hu$^{1,2}$, Xia-Ji Liu$^{2}$, and Peter D. Drummond$^{2}$}

\affiliation{$^{1}$\ Department of Physics, Renmin University of China, Beijing
100872, China \\
 $^{2}$\ ARC Centre of Excellence for Quantum-Atom Optics, Department
of Physics, University of Queensland, Brisbane, Queensland 4072, Australia}

\date{\today{}}

\begin{abstract}
We theoretically analyze a single vortex in a spin polarized 3D trapped
atomic Fermi gas near a broad Feshbach resonance. Above a critical
polarization the Andreev-like bound states inside the core become
occupied by the majority spin component. As a result, the local density
difference at the core center suddenly rises at low temperatures.
This provides a way to visualize the lowest bound state using phase-contrast
imaging. As the polarization increases, the core expands gradually
and the energy of the lowest bound state decreases. 
\end{abstract}


\maketitle
The achievement of superfluidity in trapped ultra-cold atomic $^{6}$Li
gases is a landmark advance in the history of physics \cite{MIT2005}.
This is attained by utilizing a broad Feshbach resonance, which is
used to tune the inter-atomic interactions. By changing the inverse
scattering length $a_{s}$ continuously from negative to positive
values, a two-component Fermi gas with \emph{equal} spin populations
has a ground state which crosses smoothly from Bardeen-Cooper-Schrieffer
(BCS) superfluidity to a Bose-Einstein condensate (BEC) of tightly
bound pairs. Of particular interest is the unitarity regime near resonance,
where the scattering length diverges ($1/a_{s}\simeq0$). Since the
inter-particle spacing is the only relevant length scale, the Fermi
gas exhibits a universal behavior \cite{ho2004}.

Quantized vortices are a clear-cut confirmation of superfluidity,
and were demonstrated experimentally by Zwierlein \textit{et al.}
\cite{MIT2005}. The equilibrium properties of vortices in a symmetric
Fermi superfluid at crossover have been the subject of intense theoretical
studies \cite{gygi,nygaard,bulgac,machida,mizushima,levin,ho2006}.
The Andreev-like bound states, which are the fermionic quasiparticle
excitations localized in the core, have been widely discussed \cite{gygi,machida,levin,ho2006}.
These bound states are found to play a key role in the structure of
vortices.

Most recently, Fermi gases with \emph{unequal} spin populations have
been the subject of considerable experimental \cite{MIT2006,rice}
and theoretical interest \cite{liu,sarma,bedaque,fflo,others,phasediagram,trap}.
The presence of spin polarization leads to exotic forms of pairing,
such as breached pairing \cite{liu} or Sarma superfluidity \cite{sarma},
phase separation \cite{bedaque}, and spatially modulated Fulde-Ferrell-Larkin-Ovchinnikov
(FFLO) states \cite{fflo}. An agreement on the \emph{true} ground
state of polarized fermionic superfluidity is yet to be reached. However,
three recent measurements on the density profiles of polarized $^{6}$Li
gases \cite{MIT2006,rice}, near a Feshbach resonance, indicate a
paired superfluid core surrounded by the excess unpaired fermions
consistent with a picture of phase separation.

Combining spin polarization with a vortex may help to resolve the
issue of the nature of polarized fermion pairing. It is natural to
ask how unequal spin populations affect the vortex structure, and
how vortex bound states evolve as the polarization increases. This
issue arises in the context of pairing and superfluidity in many fields
of physics \cite{rmp}. It is highly relevant to the condensed matter
community, where polarized superfluidity is created by applying a
magnetic field. There is now strong experimental evidence for the
existence of FFLO states in the heavy fermion superconductor CeCoIn$_{5}$
under high fields \cite{cecoin5}. Strongly interacting polarized
Fermi gases have also been under close scrutiny in nuclear matter
\cite{sedrakian}, neutron stars \cite{sedrakian}, and high density
quark matter \cite{alford,rmp}, where the spin polarization is created
by differences between chemical potentials and/or by mass differences
between fermions that form pairs. Polarized vortices of color superfluidity
in rotating neutron stars are a possible mechanism for observed glitches
in pulsar timing \cite{rmp}.

Here we investigate the properties of a singly quantized vortex in
polarized atomic Fermi gases at unitarity, in a cylindrically symmetric
trap. Our main results are:

(\textbf{A}) We clarify the density profiles of both spin components
as a function of polarization. In addition to phase separation, the
vortex core suddenly accommodates the excess majority fermions above
a critical polarization or a critical chemical potential difference,
resulting in a rapid rise of the local density difference inside the
core.

(\textbf{B}) The local fermionic density of states explains the sudden
appearance of an unpaired core of excess majority atoms at the vortex
center. The Andreev-like bound states in the core are occupied when
the critical chemical potential difference equals the lowest available
energy. \emph{This provides a clear visualization of vortex bound
states using phase-contrast imaging} \cite{Shin}.

(\textbf{C}) With increasing polarization, the vortex core expands
while the lowest bound state energy decreases.

The above results are obtained by numerically solving the mean-field
Bogoliubov-de Gennes (BdG) equations in a fully self-consistent fashion~\cite{gygi,BdG},
assuming a pairing order parameter that preserves the cylindrical
and axially translational symmetries. Symmetry breaking is also possible,
\textit{i.e.}, the order parameter may distort cylindrically. This
scenario merits further study.

Fermi gases of $^{6}$Li atoms near a broad Feshbach resonance are
well characterized using a single channel model \cite{sc}. The BdG
equations describing the quasiparticle wave functions $u_{\eta}\left(\mathbf{r}\right)$
and $v_{\eta}\left(\mathbf{r}\right)$, with excitation energies $E_{\eta}$
read \cite{gygi}: \begin{equation}
\left[\begin{array}{cc}
{\cal H}_{0}-\mu_{\uparrow} & \Delta(\mathbf{r})\\
\Delta^{*}(\mathbf{r}) & -{\cal H}_{0}+\mu_{\downarrow}\end{array}\right]\left[\begin{array}{c}
u_{\eta}\left(\mathbf{r}\right)\\
v_{\eta}\left(\mathbf{r}\right)\end{array}\right]=E_{\eta}\left[\begin{array}{c}
u_{\eta}\left(\mathbf{r}\right)\\
v_{\eta}\left(\mathbf{r}\right)\end{array}\right],\label{BdG}\end{equation}
 where ${\cal H}_{0}=-\hbar^{2}\mathbf{\nabla}^{2}/2m+V_{ext}\left(\mathbf{r}\right)$,
and $V_{ext}\left(\mathbf{r}\right)=m\omega^{2}\left(x^{2}+y^{2}\right)/2$
is the transverse trapping potential. Along the $z$ axis we instead
assume free motion over a length $L$. To account for the unequal
spin population $N_{\sigma}$ for $\sigma=\uparrow,\downarrow$, the
chemical potentials are shifted as $\mu_{\uparrow,\downarrow}=\mu\pm\delta\mu$,
leading to different quasiparticle wave functions for the two components.
However, there is a symmetry of the BdG equations under the replacement
$u_{\eta\downarrow}^{*}\left(\mathbf{r}\right)\rightarrow v_{\eta\uparrow}\left(\mathbf{r}\right)$,
$v_{\eta\downarrow}^{*}\left(\mathbf{r}\right)\rightarrow-u_{\eta\uparrow}\left(\mathbf{r}\right)$,
$E_{\eta\downarrow}\rightarrow-E_{\eta\uparrow}$. We can thus retain
only $u_{\eta\uparrow}\left(\mathbf{r}\right)$ and $v_{\eta\uparrow}\left(\mathbf{r}\right)$
in Eq. (\ref{BdG}), and keep solutions with both positive and negative
energies.

The order parameter $\Delta(\mathbf{r})$ and the chemical potentials
$\mu_{\uparrow,\downarrow}$ are determined by self-consistency equations
for the gap, $\Delta(\mathbf{r})=g\sum_{\eta}u_{\eta}\left(\mathbf{r}\right)v_{\eta}^{*}\left(\mathbf{r}\right)f\left(E_{\eta}\right)$,
and the particle density of each component: $n_{\uparrow}\left(\mathbf{r}\right)=\sum_{\eta}\left|u_{\eta}\left(\mathbf{r}\right)\right|^{2}f\left(E_{\eta}\right)$
and $n_{\downarrow}\left(\mathbf{r}\right)=\sum_{\eta}\left|v_{\eta}\left(\mathbf{r}\right)\right|^{2}f\left(-E_{\eta}\right)$.
These must be constrained so that $\int d\mathbf{r}n_{\sigma}\left(\mathbf{r}\right)=N_{\sigma}$,
where $f\left(x\right)=1/\left(e^{x/k_{B}T}+1\right)$ is the Fermi
distribution function, and $g$ ($<0$) is the bare coupling constant,
which is related to the $s$-wave scattering length via the regularization
prescription: $\left(4\pi\hbar^{2}a_{s}/m\right)^{-1}=1/g+\sum_{\mathbf{k}}1/2\epsilon_{\mathbf{k}}$.

We solve these equations via a \emph{hybrid} procedure, by introducing
a high energy cut-off $E_{c}$ above which we use a local density
approximation (LDA) for high-lying excitation levels. The standard
regularization prescription then yields an effective coupling constant
through the self-consistency equation $\Delta(\mathbf{r})=g_{eff}\left(\mathbf{r}\right)\sum'_{\eta}u_{\eta}\left(\mathbf{r}\right)v_{\eta}^{*}\left(\mathbf{r}\right)f\left(E_{\eta}\right)$,
where the cut-off summation $\sum'_{\eta}$ is now restricted to $\left|E_{\eta}\right|\leq E_{c}$.
Further details of this will be given elsewhere. A clear limitation
of the procedure is the use of mean-field factorizations implicit
in the BdG equations. From earlier work, we expect this to neglect
quantum fluctuations that alter the ground-state energy, while remaining
qualitatively correct \cite{hld}.

Below the cut-off, we solve the BdG equations by working in cylindrical
coordinates $\left(\rho,\varphi,z\right)$ and taking $\Delta(\mathbf{r})=\Delta(\rho)e^{-i\varphi}$
for a singly quantized vortex. Assuming periodic boundary conditions
at $z=\pm L/2$, we write, for the normalized modes, $u_{\eta}\left(\mathbf{r}\right)=u_{nmk_{z}}\left(\rho\right)e^{im\varphi}e^{ik_{z}z}/\sqrt{2\pi L}$
and $v_{\eta}\left(\mathbf{r}\right)=v_{nmk_{z}}\left(\rho\right)e^{i\left(m+1\right)\varphi}e^{ik_{z}z}/\sqrt{2\pi L}$
with $k_{z}=2\pi l/L$. As a consequence, the BdG equations decouple
into different $m$ and $l$ sectors \cite{gygi}. Expanding the radial
functions $u_{nmk_{z}}\left(\rho\right)$ and $v_{nmk_{z}}\left(\rho\right)$
in a basis set of 2D harmonic oscillators, we then solve a matrix
eigenvalue problem in each sector.

%
\begin{figure}
\begin{centering}\includegraphics[clip,width=0.45\textwidth]{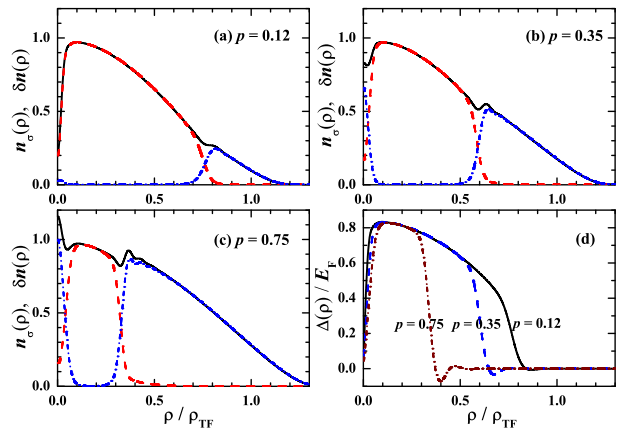}\par\end{centering}

\caption{(Color online). Density profiles of the majority ($\uparrow$-state,
solid lines) and minority ($\downarrow$-state, dashed lines) components
at $T=0.05T_{F}$ for three typical value of polarization: $p=0.12$
(a), $p=0.35$ (b), and $p=0.75$ (c). Density differences are also
plotted in dotted-dashed lines. All the profiles are normalized by
$n_{\sigma},_{TF}=\left(1+\beta\right)^{-3/5}\sqrt{15\pi N\lambda/2}/\left(6\pi^{2}\right)\left(\hbar/m\omega\right)^{-3/2}$,
which is the peak density for a symmetric gas at unitarity. Panel
(d) shows the order parameter profiles. The small oscillations at
the edge are a finite size effect.}

\label{fig1} 
\end{figure}


In greater detail, we consider a gas at unitarity with the number
of total atoms in the range $N=N_{\uparrow}+N_{\downarrow}=2\times10^{3}-4\times10^{4}$.
Two characteristic scales may be defined by considering a symmetric
ideal Fermi gas at zero temperature. In the LDA analysis this leads
to a Thomas-Fermi (TF) radius $\rho_{TF}^{0}=\left(15\pi N\lambda/2\right)^{1/6}\sqrt{\hbar/m\omega}$,
and a Fermi energy $E_{F}=\left(15\pi N\lambda/16\right)^{1/3}\hbar\omega\equiv k_{B}T_{F}$,
where we define $\lambda=L/\rho_{TF}^{0}$ as the aspect ratio of
the trap. Throughout this Letter, we calculate results at the Feshbach
resonance with $1/a_{s}=0$ and use $\lambda=1$ and $E_{c}\simeq2E_{F}$.
We also considered coupling constants in the BCS regime but observed
no significant changes. Dimensionality effects will be treated elsewhere.

Numerical accuracy was checked by increasing $E_{c}$ up to $4E_{F}$.
Due to the high accuracy of our hybrid cut-off procedure, the results
were found to be essentially independent of the cut-off energy. We
note finally that, for a symmetric gas at unitarity, universality
implies a TF radius of $\rho_{TF}=\left(1+\beta\right)^{3/10}\rho_{TF}^{0}$,
a chemical potential $\mu=\left(1+\beta\right)^{3/5}E_{F}$, and a
maximum order parameter $\Delta_{0}=8\left(1+\beta\right)^{3/5}E_{F}/e^{2}$
\cite{ho2004}, where BCS theory predicts the universal parameter
$\beta\simeq-0.41$.

We present in Figs. 1a, 1b and 1c the density profile of each component,
as well as the density difference $\delta n\left(\mathbf{r}\right)=$
$n_{\uparrow}\left(\mathbf{r}\right)-n_{\downarrow}\left(\mathbf{r}\right)$,
for several polarizations $p=\left(N_{\uparrow}-N_{\downarrow}\right)/N$
at $T=0.05T_{F}$ and $N=10^{4}$. Because of the uniform distribution
along the $z$ axis, these profiles are linked to the experimentally
observed column densities in the axial direction. Apart from the apparent
phase separation at the edge, the most salient feature of the figures
is the development of a polarized normal shell inside the vortex core
above a certain polarization. This is clearly visible as a prominent
peak in the density difference, of width about $0.05\rho_{TF}$. This
is observable in the column integrated density difference, which is
directly measurable by phase-contrast imaging \cite{MIT2006,Shin}.

%
\begin{figure}
\begin{centering}\includegraphics[clip,width=0.45\textwidth]{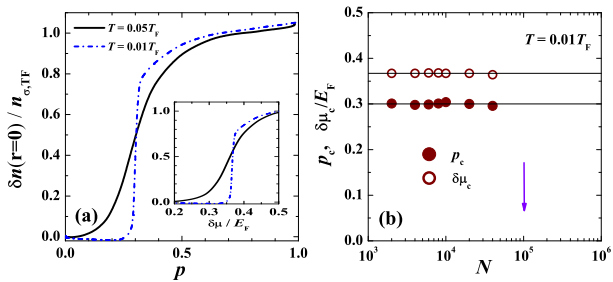}\par\end{centering}

\caption{(Color online). Left panel: centre density difference as a function
of polarization at $N=10^{4}$. Inset shows the dependence on the
chemical potential difference. Right panel: critical polarization
and critical chemical difference as a function of the number of total
particles.}

\label{fig2} 
\end{figure}


The onset of a polarized normal shell at the core center is demonstrated
by the central density difference as a function of the polarization.
This is shown in Fig. 2a, which represents the most important result
of this Letter. At a sufficiently low temperature, \textit{i.e.},
$T=0.01T_{F}$, a sudden rise of the center density difference appears
at a critical polarization $p_{c}\simeq0.30$. The critical chemical
potential difference is $\delta\mu_{c}\simeq0.36E_{F}\sim\Delta_{0}^{2}/2E_{F}$,
with a transition width of around $k_{B}T$. This transition is therefore
much smoother at finite temperature. The critical polarization is
nearly independent of the overall number of atoms $N$, as shown in
Fig. 2b for $N$ up to $4\times10^{4}$. We therefore expect that
this will apply to current experiments, where the typical number of
atoms is around $10^{5}$, and would survive even in the thermodynamic
limit.

The appearance of this intriguing shell structure is closely related
to the Andreev-like bound states inside the core. In the BCS regime,
these states are formed by the spatial variation of the order parameter
around the center (see, \textit{i.e.}, Fig. 1d), analogous to a potential
well for quasiparticles, of depth $\Delta_{0}$ and of radius equal
to the coherence length $\xi=\hbar v_{F}/\Delta_{0}$. Hence, the
confinement of the well gives rise to discrete bound levels with spacing
of order $\hbar^{2}/m\xi^{2}=\Delta_{0}^{2}/2E_{F}$ \cite{gygi}.
This qualitative picture persists in the strongly interacting unitarity
limit \cite{ho2006}.

%
\begin{figure}
\begin{centering}\includegraphics[clip,width=8.5cm]{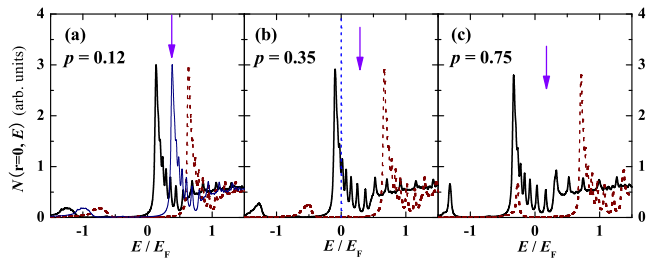}\par\end{centering}

\caption{(Color online). Local fermionic density of states of spin up (solid
lines) and spin down (dashed lines) components inside the vortex core
at $N=10^{4}$ and $T=0.05T_{F}$. The thin line in (a) shows the
LDOS at $p=0$. Arrows points to the position at the effective energy
of the lowest bound state.}

\label{fig3} 
\end{figure}


To provide an intuitive explanation of our results, we calculate the
local density of states (LDOS), \begin{eqnarray}
N_{\uparrow}\left(\mathbf{r},E\right) & = & \sum\nolimits _{\eta}\left|u_{\eta}\left(\mathbf{r}\right)\right|^{2}\delta\left(E-E_{\eta}\right),\nonumber \\
N_{\downarrow}\left(\mathbf{r},E\right) & = & \sum\nolimits _{\eta}\left|v_{\eta}\left(\mathbf{r}\right)\right|^{2}\delta\left(E+E_{\eta}\right).\end{eqnarray}
 At low temperature, when integrated over negative energy, this leads
to the density profiles $n_{\sigma}\left(\mathbf{r}\right)$. In Fig.
3 we show how the LDOS inside the core evolves with increasing the
polarization. A small spectral broadening of about $0.01E_{F}$ has
been used to regularize the delta function. Without any polarization
the LDOS of the two components coincides, leading to a sharp peak
located at positive energy $E_{bs}^{0}\simeq\Delta_{0}^{2}/2E_{F}$,
associated with the lowest Andreev-like bound state.

In the presence of spin-polarization the peak in the density of states
shifts in different directions for the two components. To a good approximation,
the energy separation between the two peaks at the vortex center equals
$2\delta\mu$. Thus, in the general case of a nonzero polarization
one may define an effective energy of the lowest bound state, $E_{bs}$,
as the midpoint of these two peaks located at $E_{bs}\mp\delta\mu$.
Therefore, a net density difference results precisely when the peak
in $N_{\uparrow}\left(\mathbf{r=0},E\right)$ crosses zero energy
\textit{i.e.}, $\delta\mu=E_{bs}$. This results in a bound state
for the majority spin component, which explains why a polarized normal
shell emerges above a critical population chemical potential $\delta\mu_{c}\sim E_{bs}^{0}\simeq\Delta_{0}^{2}/2E_{F}$.

Thus, the integrated column density difference is an indicator of
the lowest vortex bound state, and a measurement of the critical polarization
$p_{c}$ gives its energy.


We now consider the dependence of the vortex core size on the polarization.
We extract the core size from the superfluid density $n_{s}\left(\mathbf{r}\right)$,
defined as a ratio of the current density $\mathbf{j}\left(\mathbf{r}\right)=n_{s}\left(\mathbf{r}\right)\mathbf{v}_{s}$
to the superfluid velocity $\mathbf{v}_{s}=\left(\hbar/2m\rho\right)\mathbf{\hat{\varphi}}$
\cite{ho2006}, where, since our normal fluid solutions are non-rotating:
\begin{equation}
\mathbf{j}\left(\mathbf{r}\right)=\frac{i\hbar}{m\rho}\sum_{\eta}\left[u_{\eta}^{*}\partial_{\varphi}u_{\eta}f\left(E_{\eta}\right)+v_{\eta}\partial_{\varphi}v_{\eta}^{*}f\left(-E_{\eta}\right)\right]{\bf \hat{\varphi}}.\end{equation}
 The resulting superfluid density profiles are plotted in the inset
of Fig. 4a. The core size may be quantified as the distance from the
vortex core at which the superfluid density is 90\% of its maximum
value, namely, $\xi_{90}$. From Fig. 4a, the core size increases
gradually with increasing polarization, and almost doubles at large
polarization.

%
\begin{figure}
\begin{centering}\includegraphics[clip,width=8.5cm]{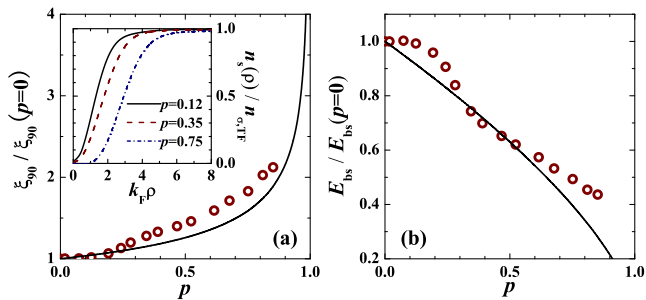}\par\end{centering}

\caption{(Color online). Vortex core size (a) and the effective energy of
the lowest bound state (b) as a function of polarization at $N=10^{4}$
and $T=0.05T_{F}$. The core size $\xi_{90}$ at $p=0$ is about $2.5k_{F}^{-1}$,
where $k_{F}$ non-interacting Fermi wavelength at center. Solid lines
are the scaling relations as described in the text. Inset shows superfluidity
density profiles.}

\label{fig4} 
\end{figure}

To explain this, note that while a phase separation occurs at \emph{any}
nonzero polarization, only the unpolarized superfluid part can form
a vortex. Thus, the vortex core should expand with a scaling of $\xi_{90}\propto\left(2N_{\downarrow}\right)^{-1/3}\propto\left(1-p\right)^{-1/3}$
\cite{ho2006}. In Fig 4a this scaling is plotted by a solid line,
which fits well with our numerical results. Accordingly, one may suspect
that the energy of the lowest bound state will decrease as $E_{bs}\propto1/$
$\xi_{90}^{2}\propto\left(1-p\right)^{2/3}$. This is consistent with
the effective energy of the lowest bound state shown in Fig. 4b. We
expect a phase separation into \emph{multiple} vortex cores in a vortex
lattice, as in current non-polarized experiments \cite{MIT2005}.

We have considered an aspect ratio $\lambda=1$, which is closer to
the MIT experimental setup \cite{MIT2006} than the Rice experiment
(which has $\lambda=50$ \cite{rice}). In the opposite limit of $\lambda\ll1$,
an interesting aspect of dimensionality would arise. Due to strong
phase fluctuations, this quasi-2D geometry would favor the spontaneous
formation of vortices at finite temperature \cite{KBT}. As a result,
a lattice of vortex-anti-vortex pairs \emph{without} phase separation
may emerge as the ground state. In such a configuration, the spin
polarization would be sustained by a polarized normal shell \emph{inside}
the vortex cores, analogous to a type-II superconductor in a magnetic
field.

In conclusion, we have analyzed vortex structures in a polarized Fermi
gas at unitarity. The lowest bound state will be visible via phase-contrast
imaging, together with a quantum phase transition at a critical spin
polarization.

This work is supported by the Australian Research Council Center of
Excellence and by the National Science Foundation of China Grant
No. NSFC-10574080 and the National Fundamental Research Program under
Grant No. 2006CB921404. One of us (PDD) acknowledges useful discussions
with Yong-Il Shin.

\end{document}